# Enhanced Load Balancing Approach to Avoid Deadlocks in Cloud

Rashmi. K. S
Post Graduate Programme, Computer Science and Engineering, Department of Information Science and Engineering, Dayananda Sagar College of Engineering, Bangalore, India

Suma. V
Department of Information Science and Engineering, Dayananda Sagar College of Engineering, Bangalore, India

Vaidehi. M
Research and Industry Incubation Center, Dayananda Sagar Institutions, Bangalore, India

## ABSTRACT
The state-of-art of the technology focuses on data processing to deal with massive amount of data. Cloud computing is an emerging technology, which enables one to accomplish the aforementioned objective, leading towards improved business performance. It comprises of users requesting for the services of diverse applications from various distributed virtual servers. The cloud should provide resources on demand to its clients with high availability, scalability and with reduced cost. Load balancing is one of the essential factors to enhance the working performance of the cloud service provider. Since, cloud has inherited characteristic of distributed computing and virtualization there is a possibility of occurrence of deadlock. Hence, in this paper, a load balancing algorithm has been proposed to avoid deadlocks among the Virtual Machines (VMs) while processing the requests received from the users by VM migration. Further, this paper also provides the anticipated results with the implementation of the proposed algorithm. The deadlock avoidance enhances the number of jobs to be serviced by cloud service provider and thereby improving working performance and the business of the cloud service provider.

## General Terms
Cloud Computing, Load Balancing, Virtual Machines (VMs).

## Keywords
Deadlock Avoidance, VM Migration, Respone Time, Hop Time, Wait Time.

## 1. INTRODUCTION
The success of IT organizations lies in acquiring the resources on demand. Cloud computing is a promising technology to provide on demand services according to the clients requirements within a stipulated time. Further, the cloud computing environment provides the users for accessing the shared pool of distributed resources. Cloud is a pay- go model where the consumers pay for the resources utilized instantly, which necessitates having highly available resources to service the requests on demand. Hence, the management of resources becomes a complex job from the business perspective of the cloud service provider. Further, there can be a scenario where in the cloud service provider's datacenter will be hosting less number of Virtual Machines (VMs) compared to the number of jobs arrived for availing the service. In such a situation, similar types of jobs will be competing to acquire the same VM at the same time leading to a deadlock. Further, the deadlock problem leads to the degradation of working performance as well as the business performance of the cloud service provider. Henceforth, an efficient load balancing technique is required to distribute the load to avoid deadlocks.

To achieve the above mentioned objective a load balancing algorithm that supports migration has been proposed in this paper. In the cloud computing environment the load refers to the number of requests that has to be serviced by VMs that are available in cloud. The proposed algorithm avoids the deadlock by providing the resources on demand resulting in increased number of job executions. Henceforth, the business performance of the cloud service provider is improved by reducing the rejection in the number of jobs submitted [7]. In order to implement the proposed technique hop time and wait time may be considered. Hop time is the duration involved in migration of the job from the overloaded VM to the underutilized VM for providing the service. Wait time is the time after which the VMs become available to service the request.

The paper is organized as follows. The following section details the related work followed by the design model in Section 3. Section 4 discusses the existing and the proposed algorithm. Section 5 gives the simulation setup and expected results. Finally, the paper is concluded in Section 6 followed by references in Section 7.

## 2. RELATED WORK
Cloud Computing is the recent emergence of advanced technology in the IT industries leading towards opening of several research avenues in the domain. Shu-Ching Wang et al. have proposed a Load Balancing in a three –level cloud computing network, by using a scheduling algorithm which combines the features of Opportunistic Load Balancing (OLB) and Load Balance Min-Min (LBMM) which can utilize better executing efficiency and maintain load balancing of the system. The objective is to select a node based for executing the complicated tasks that needs large-scale computation. The scheduling algorithm proposed in this paper is not dynamic and also there is an overhead involved in the selection of the node [1].

Vlad Nae et al. have implemented Cost-Efficient Hosting and Load Balancing of Massively Multiplayer Online Games with the objective to reduce hosting costs and to achieve resource allocation by load distribution so that QoS constraint is satisfied at all times. Further, the authors have presented on-demand prediction-based resource allocation and load balancing method for real-time MMOGs. The load balancing algorithm fails to optimize the distribution of load by considering both cost and resource characteristics [2].

Hao Liu et al. have proposed LBVS: A Load Balancing Strategy for Virtual Storage to provide a large scale net data storage model and Storage as a Service model based on cloud. Further the storage virtualization is achieved using three layers architecture with two load balancing modules to balance the load. The strategy implemented in this paper is limited to the cloud service providers providing Storage as a Service (SaaS) [3].

HUIWen et al. have recommended an Effective Load Balancing for Cloud-based Multimedia System called CMLB to allocate and schedule resources for different user requests in a very reasonable way by providing a cloud-based framework for multimedia applications, which provides a good solution to the inherent issues of multimedia applications, such as computational complexity and multimedia QoS provisioning. This approach considers the network conditions to distribute, which is an overhead. [4].

Wenhong Tian et al. have introduced a dynamic and integrated load balancing scheduling algorithm (DAIRS) for cloud datacenters, with the objective to develop an integrated measurement for the total imbalance level of a Cloud datacenter as well as the average imbalance level of each server. The algorithm is time consuming during the resource allocation, as it sorts the physical servers in an ascending order of their utilization [5].

## 3. DESIGN MODEL

It is evident from the progress of our survey that none of the load balancing techniques are efficient in avoiding deadlocks among the VMs. Henceforth, a load balancing algorithm to avoid deadlock by incorporating the migration has been proposed. Figure 1. depicts the design of the cloud architecture for this approach.

According to this design various users submit their diverse applications to the cloud service provider through a communication channel. The Cloud Manager in the cloud service provider's datacenter is the prime entity to distribute the execution load among all the VMs by keeping track of the status of the VM. The Cloud Manager maintains a data structure containing the VM ID, Job ID of the jobs that has to be allocated to the corresponding VM and VM Status to keep track of the load distribution. The VM Status represents the percentage of utilization. The Cloud Manager allocates the resources and distributes the load as per the data structure. The Cloud Manager analyzes the VM status routinely to distribute the execution load evenly. In course of processing, if any VM is overloaded then the jobs are migrated to the VM which is underutilized by tracking the data structure. If there are more than one available VM then the assignment is based on the least hop time. On completion of the execution, the Cloud Manager automatically updates the data structure. Table 1 illustrates the sample data structure maintained by the Cloud Manger.

Further, the load balancing approach can be represented using a mathematical model. Let the graph be G = (V,E), where V is the disjoint vertex set represented as V= v1 U v2 in which v1 represents the set of VMs in the datacenter and v2 represents the set of jobs received and E is the mapping between the two sets of vertices. Figure 2. depicts the mathematical model.

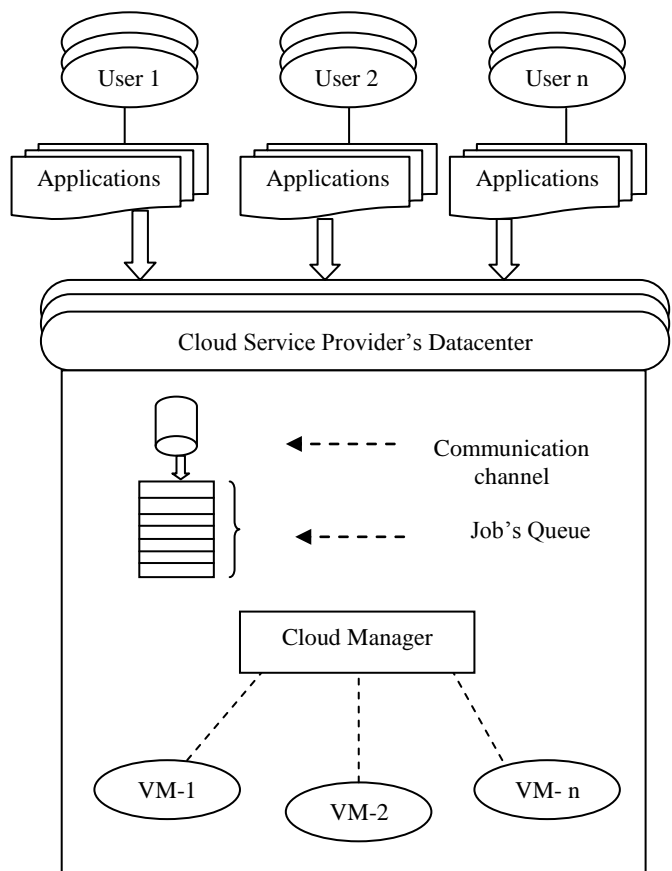

**Figure 1. Cloud Architecture for Load Balancing**

**Table 1. Cloud Manager Data Structure**

| Serial No. | Job ID | VM ID | VM Status (%) |
|---|---|---|---|
| 1 | $Job_1, Job_2$ | $VM_1$ | 20 |
| 2 | $Job_2, Job_1, Job_n$ | $VM_2$ | 100 |
| 3 | $Job_3$ | $VM_3$ | 50 |
| 4 | $Job_4, Job_2$ | $VM_4$ | 10 |
| . | . | . | . |
| . | . | . | . |
| . | . | . | . |
| n | $Job_4, Job_n$ | $VM_{n-1}$ | 30 |

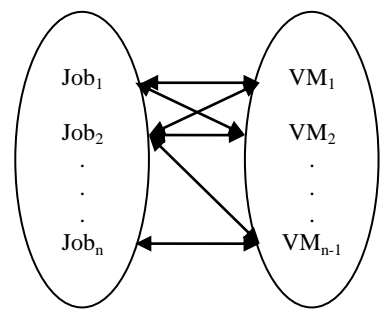

**Figure 2. Mathematical Model**

From Table 1. it can be inferred that the status of $VM_2$ is 100% which means it is completely utilized and Figure 2. infers that there are more than two jobs arriving to acquire the $VM_2$, which will be overloaded at a same point of time causing a deadlock. Henceforth, the proposed load balancing algorithm identifies the underutilized VM and migrates the load to them for avoiding the deadlock.

## 4. ALGORITHM

Algorithm facilitates the transformation of mathematical model to implementation model. The existing load balancing algorithm and the proposed algorithms are given below.

### 4.1 Existing Algorithm

**Step 1.** The Load Balancer maintains an index table of VMs and the number of requests currently allocated to the VM. At the start all VM's have 0 allocations.

**Step 2.** When a request to allocate a new VM from the DataCenterController arrives, it parses the table and identifies the least loaded VM. If there are more than one, the first identified is selected.

**Step 3.** The Load Balancer returns the VM ID to the DataCenterController.

**Step 4.** The DataCenterController sends the request to the VM identified by that id.

**Step 5.** DataCenterController notifies the Load Balancer of the new allocation.

**Step 6.** The Load Balancer updates the allocation table incrementing the allocations count for that VM.

**Step 7.** When the VM finishes processing the request, and the DataCenterController receives the response cloudlet, it notifies the Load Balancer of the VM de-allocation.

**Step 8.** The Load Balancer updates the allocation table by decrementing the allocation count for the VM by one.

**Step 9.** Continue from step 2.

In the existing algorithm there exists a communication between the Load Balancer and the DataCenterController for updating the index table leading to an overhead. Further, this overhead causes delay in providing response to the arrived requests.

### 4.2 Proposed Algorithm: Enhanced Load Balancing Algorithm using Efficient Cloud Management System

**Step1.** Initially VM status will be 0 as all the VMs are available. Cloud Manager in the datacenter maintains a data structure comprising of the Job ID, VM ID and VM Status.

**Step2.** When there is a queue of requests, the cloud manager parses the data structure for allocation to identify the least utilized VM. If availability of VMs is more then, the VM with least hop time is considered.

**Step3.** The Cloud Manager updates the data structure automatically after allocation.

**Step4.** The Cloud Manager periodically monitors the status of the VMs for the distribution of the load, if an overloaded VM is found, and then the cloud manager migrates the load of the overloaded VM to the underutilized VM.

**Step5.** The decision of selecting the underutilized VM will be based on the hop time. The VM with least hop time is considered.

**Step6.** The Cloud Manager updates the data structure by modifying the entries accordingly on a time to time basis

**Step7.** The cycle repeats from Step2.

In the proposed algorithm the Cloud Manager analyses the availability of the VMs at the time of job arrivals to update the data structure thereby having less overhead involved in maintenance of the data structure compared to the existing approach.

## 5. SIMULATION SETUP AND EXPECTED RESULTS

The simulation setup of the proposed approach will comprise the following configurations. Table 2. depicts the simulation configuration of the user. The user configuration comprises of the Job ID and Job capacity that each job will be requesting and it is assumed that all the requests have arrived at the same time.

Table 3. depicts the datacenter configuration, which comprises the VM ID and the VM capacity.

**Table 2. User Configuration**

| Job ID | Job Capacity |
|---|---|
| J1 | 1000 |
| J2 | 10000 |
| J3 | 1000 |
| J4 | 100 |
| J5 | 10000 |
| J6 | 100000 |

**Table 3. Datacenter Configuration**

| VM ID | VM Capacity |
|---|---|
| V1 | 100 |
| V2 | 1000 |
| V3 | 1000 |
| V4 | 100000 |
| V5 | 10000 |

It is inferred from Table 2. and Table 3. that the number of VMs available is less compared to the number of jobs that have arrived. Henceforth, there will be at least two jobs competing to acquire the same VM leading to an occurrence of deadlock. By considering the aforementioned simulation configurations the existing algorithm has been evaluated. Table 4. depicts its result in terms of response time.

**Table 4. Response Time obtained for Existing Algorithm**

| Job ID | Response Time (ms) |
|---|---|
| J1 | 528.99 |
| J2 | 1031.557 |
| J3 | 330.206 |
| J4 | 81.096 |
| J5 | 1029.967 |
| J6 | 1034.52 |

It can be inferred from Table 4. that the occurrence of deadlock results in high response time. Figure 3. depicts the response time graph of each job obtained from the evaluation of the existing load balancing algorithm.

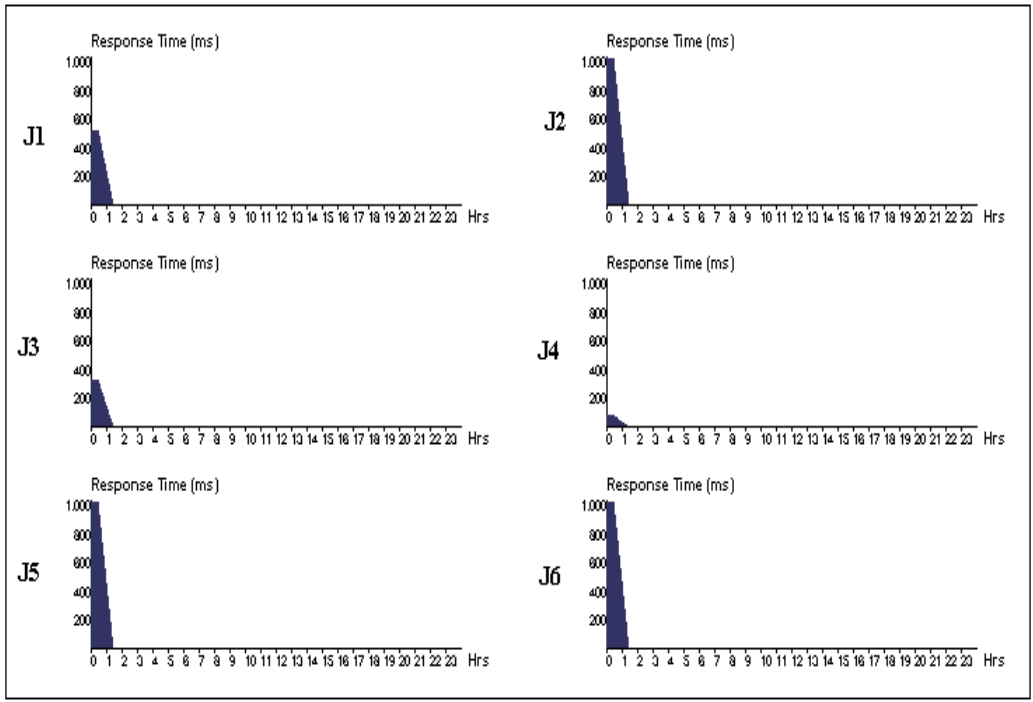

**Figure 3. Graph of Response Time using Existing Algorithm**

The increased response time to service the request, which arises due to the occurrence of deadlock, will increase the job rejection rate. On increased job rejections, the business performance of the cloud service provider will deteriorate. It is worth to note that the service provider should respond early to the jobs arrived in order to have a good business.

Henceforth, the proposed load balancing algorithm must yield less response time for the same simulation configuration. The response time obtained using the proposed approach must be at least 30 %-50% less compared to the one obtained by evaluating the existing approach. Henceforth, Table 5. depicts the expected outcome of the proposed load balancing approach.

**Table 5. Expected Response Time to be Obtained from Evaluating the Proposed Algorithm**

| Job ID | Response Time (ms) |
|---|---|
| J1 | 250.33 |
| J2 | 489.68 |
| J3 | 250.33 |
| J4 | 53.67 |
| J5 | 489.679 |
| J6 | 504.99 |

From Table 5. it is analyzed that with less response time the job rejections will be reduced thereby enhances the business performance of the cloud service provider. Figure 4. shows the graphical comparison between the existing load balancing algorithm and proposed approach .

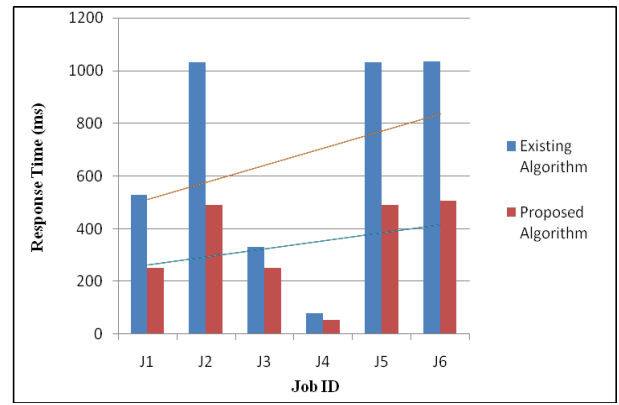

**Figure 4. Graphical comparison of existing algorithm and the proposed algorithm**

It is observed from Figure 4. that the implementation of proposed approach yields less response time compared to the existing approach . Thus, less response time reduces job rejections and accelerates the business performance.

Further, the efficiency of the proposed algorithm to VM migration may be evaluated by comparing the hop time to move from the overloaded VM to the underutilized VM and the wait time of the VM to become available to service the request.

The wait time will be considered when the time to wait for the VM to become available is less compared to the hop time.

As the overhead is involved by the existing approach for keeping track of the available resources and for updating of the VM status time to time, the proposed approach overcomes this by analyzing the availability of VMs on a time to time basis.

The scope of this paper limits to analyzing the efficiency of avoiding the deadlocks using the enhanced load balancing approach.

## 6. CONCLUSION

Cloud computing is a promising technology, which is a pay-go model that provides the required resources to its clients. Since, virtualization is one of the core characteristics of cloud computing it is possible to virtualize the factors that modulates business performance such as IT resources, hardware, software and operating system in the cloud-computing platform. Further, the cloud having the characteristic of distributed computing there can be chances of deadlocks occurring during the resource allocation process.

The load balancing is implemented in the cloud computing environment to provide on demand resources with high availability. But the existing load balancing approaches suffers from various overhead and also fails to avoid deadlocks when there more requests competing for the same resource at a time when there are resources available are insufficient to service the arrived requests.

The enhanced load balancing approach using the efficient cloud management system is proposed to overcome the aforementioned limitations. The evaluation of the proposed approach will be done in terms of the response time and also by considering the hop time and wait time during the migration process of the load balancing approach to avoid deadlocks.

Henceforth, the proposed work improves the business performance of the cloud service provider by achieving the aforementioned objectives.